\documentclass[final,3p,times,twocolumn]{elsarticle}

\usepackage{hyperref}
\usepackage[usenames,dvipsnames]{color}
\usepackage{amsmath}
\usepackage{amssymb}
\usepackage{amsthm}
\usepackage{array}
\usepackage{graphicx}
\usepackage{epstopdf}
\usepackage[normalem]{ulem}
\usepackage{verbatim}
\usepackage{stmaryrd}
\usepackage{lipsum}
\usepackage{mathtools}
\usepackage{xcolor}
\usepackage{diagbox}
\usepackage{tabularx}
\normalsize
\usepackage{booktabs}
\usepackage{setspace}

\usepackage{newfloat}
\makeatletter
\let\newfloat\newfloat@ltx
\makeatother
\usepackage{algorithm}
\usepackage{algpseudocode}
\usepackage{bm}

\usepackage[export]{adjustbox}

\newcommand{\T}{\mathcal{T}}

\newcommand{\bG}{\mathbf{G}}
\newcommand{\bSigma}{\mathbf{\Sigma}}
\newcommand{\bS}{\mathbf{S}}

\newcommand{\bP}{\mathbf{P}}
\newcommand{\bX}{\mathbf{X}}

\newcommand{\bO}{\mathbf{O}}
\newcommand{\bV}{\mathbf{V}}
\newcommand{\bD}{\mathbf{D}}
\newcommand{\pySCF}{\texttt{pySCF} }
\newcommand{\zgemm}{\texttt{ZGEMM} }
\newcommand{\bk}{\mathbf{k}}
\newcommand{\bq}{\mathbf{q}}
\newcommand{\bK}{\mathbf{K}}

\newcommand{\br}{\mathbf{r}}
\newcommand{\bR}{\mathbf{R}}

\newcommand{\bU}{\mathbf{U}}
\newcommand{\bH}{\mathbf{H}}
\newcommand{\bJ}{\mathbf{J}}
\newcommand{\bC}{\mathbf{C}}

\journal{Computer Physics Communications}
\begin{document}
\begin{frontmatter}

\title{Symmetry adaptation for self-consistent many-body calculations}
\author[Beijing,AnnArbor]{Xinyang Dong}
\author[AnnArbor]{Emanuel Gull}
\date{\today} 
\affiliation[Beijing]{AI for Science Institute, Beijing 100080, China}
\affiliation[AnnArbor]{Department of Physics, University of Michigan, Ann Arbor, MI 48109, USA}

\begin{abstract}
The exploitation of space group symmetries in numerical calculations of periodic crystalline solids accelerates calculations and provides physical insight.
We present results for a space-group symmetry adaptation of electronic structure calculations within the finite-temperature self-consistent GW method along with an efficient parallelization scheme on accelerators.
Our implementation employs the simultaneous diagonalization of the Dirac characters of the orbital representation.
Results show that symmetry adaptation in self-consistent many-body codes results in substantial improvements of the runtime, and that block diagonalization on top of a restriction to the irreducible wedge results in additional speedup.
\end{abstract}

\end{frontmatter}


\section{Introduction}
Space group symmetries play a critical role in understanding crystalline materials. They enable the interpretation of properties, such as optical or electronic response functions, in terms of irreducible representations, revealing information such as the dominant atomic character of orbitals or the set of allowed optical transitions \cite{Dresselhaus_2008}.
In numerical calculations, the consideration of these symmetries leads to a substantial acceleration of computer codes.

Standard electronic structure codes, such as implementations of the Hartree Fock, Density Functional Theory, or non-selfconsistent GW method, are predominantly formulated in terms of single-particle density matrices for which the symmetry adapation formalism is well established.
In contrast, modern many-body technology relies on the repeated and numerically involved calculation of objects beyond the density matrix, such as the self-consistent evaluation of  frequency-dependent propagators, vertex functions, and screened interactions. These calculations gain an even stronger advantage from considering all symmetries, including those resulting from non-symmorphic space group elements, and the repeated use of symmetrized objects due to frequency summations justify the larger initial cost of generating a fully symmetry-adapted basis.

In addition,  computational architectures such as GPUs and other accelerator platforms are able to efficiently utilize the parallelism exposed by the group structure, resulting in a substantial additional speedup.

A rigorous, general, and numerically elegant formulation of symmetry adaptation in periodic solids has been pioneered in a series of groundbreaking papers by Dovesi and collaborators \cite{Dovesi_1986,Dovesi_1998,Dovesi_1998_1} in the context of Hartree Fock and DFT theory, and was implemented in the `crystal' code \cite{Dovesi_2018,Dovesi_2014}. This paper will revisit this formalism and present results for the self-consistent GW method \cite{Hedin_1965,Yeh_22} on modern GPU architectures, illustrating the advantage of symmetry adaptation by block-diagonalization on top of a restriction to the irreducible wedge.
In addition, we present parallelization strategies and scaling results for a set of paradigmatic solids.

\section{Model and Method}
\subsection{System and Hamiltonian}
We consider a generic non-relativistic electronic structure setup \cite{Martin_2004}, $\hat{H} = \hat{H}_0 + \hat{U}$, consisting of a one-electron part $\hat{H}_0$ and two-electron Coulomb interaction $\hat{U}$. We assume translation invariance and a finite system with periodic boundary condition such that crystal momentum is conserved. In second quantization, the Hamiltonian $H$ is 
\begin{align}
    H = 
    &\sum_{\bk}\sum_{ij}\sum_{\sigma \sigma'} (H_0)^{\bk}_{i\sigma,j\sigma'} {c}^{\bk\dag}_{i\sigma} {c}^{\mathbf{k}}_{j\sigma'} \\
    &+ \frac{1}{2N_{k}}\sum_{ijkl}\sum_{\bk_{i}\bk_{j}\bk_{k}\bk_{l}}\sum_{\sigma\sigma'}U^{\bk_{i}\bk_{j} \bk_{k}\bk_{l}}_{\ i\ j\ \ k \ l} {c}^{\bk_{i}\dag}_{i\sigma} {c}^{\bk_{k}\dag}_{k\sigma'} {c}^{\bk_{l}}_{l\sigma'} {c}^{\bk_{j}}_{j\sigma} \nonumber,
\end{align}
where $\bH_0$ is the one-electron Hamiltonian, $N_k$ is the number of momentum points in the Brillouin zone and $c_{i\sigma}^{\bk (\dagger)}$ annihilate (create) an electron in a single-particle state with momentum $\bk$, spin $\sigma$ and basis state (orbital) number $i$. 
Momentum conservation of the scattering implies $\bk_i + \bk_k - \bk_j - \bk_l = \mathbf{K}$, where $\bK$ is a reciprocal lattice vector. 

\subsection{Single-particle basis states}

Single-particle states are Bloch waves characterized by a momentum label $\bk$, a spin label $\sigma$, and an orbital index $i$. In our case, they consist of linear superposition of atom-centered wave functions $g_{i}^{\bR}(\br) = g_{(x j)}^{\bR}(\br)$, where $\bR$ denotes a unit cell translation vector, $x$ enumerates the atom inside the unit cell, $j$ denotes its atomic orbital index, and $i = (xj)$ is a unit-cell orbital index enumerating all orbitals in the unit cell
\begin{align}
    &g^{\bk}_{i}(\mathbf{r}) = \sum_{\bR} g_{i}^{\bR}(\br) e^{i\bk\cdot\mathbf{R}} = \sum_{\bR} g_{(x j)}^{\bR}(\br) e^{i\bk\cdot\mathbf{R}} \, . 
\end{align}
While there is considerable freedom in the choice of basis functions and the remainder of the work will be valid for any function of this form, our implementation \cite{GREEN} employs standard atom-centered linear combination of Gaussian-type orbitals.

\section{Symmetries of solids}

\subsection{Space group symmetries}
A precise theory of symmetries in solid was established already in the 19th century \cite{Schoenflies_1891}, and the mathematical \cite{Bradley_1972} and physical \cite{Lax_1974, Dresselhaus_2008} formalism is well developed.
Computational approaches for mean field theories are also well developed \cite{Dovesi_1998, Dovesi_1998_1}, in particular the adaptation to the ``irreducible wedge''.
Rather than giving an extensive description of the theory, we recapitulate use those approach that enable an efficient generalization to many-body theories, following to a large extent the work in \cite{Dovesi_1998, Dovesi_1998_1}.
We refer the intended reader to \cite{Lax_1974} for introductory material, \cite{Bradley_1972} for a mathematical exposition, \cite{BilbaoServer} for useful tables and \cite{Dovesi_1998} for practical implementation in computer codes.

We consider the space-group symmetries and time-reversal. Space group symmetries consist of translations, point-group operations such as rotations or reflections, or combinations of the two. Only 230 such groups exist in three dimension.
We denote symmetry elements as $\hat{\alpha} = \{\alpha|v(\alpha)\}$, in which $\alpha$ is a point group operation and $v(\alpha)$ is a translation. Groups containing elements for which $v(\alpha)$ is neither zero nor a lattice translation are called non-symmorphic (there are 157 such groups); the remaining groups are called symmorphic \cite{Lax_1974}.

\subsection{Symmetry operations on orbitals}
The action of a symmetry operation $\hat{\alpha}$ on orbital $j$ of atom $x$ at momentum $\bk$ can be written as \cite{Dovesi_1998, aims_2022}
\begin{align}
    \hat{\alpha} g^{\bk}_{(xj)}(\mathbf{r}) 
    &= \sum_{\bR} e^{i\bk\cdot\mathbf{R}} [\hat{\alpha} g_{(x j)}^{\bR}(\br)] \nonumber \\ 
    &= \exp[-i \tilde{\bk} \cdot \mathbf{v}_{x}^{\alpha}] \times [\mathbf{O}({\alpha}) g^{\tilde{\bk}}_{(xj)}(\mathbf{r})] \, ,
    \label{eq:basis_rot}
\end{align}
where $\alpha\bk = \tilde{\bk} + \bK$, with $\bK$ the lattice vector that shifts $\alpha\bk$ back to the first Brillouin zone, $\mathbf{v}_{x}^{\alpha}$ the translation vector between rotated and original atom, and $\bO({\alpha})$ is the representation matrix of operation ${\alpha}$ in the basis spanned by $g_{(xj)}(\br)$. 

To obtain the explicit form of $\bO^{\alpha}$, we consider the angular part of the basis functions. In standard Gaussian-type bases, the angular part of atomic orbitals are real spherical harmonics $Y_{lm}$, which are linear combinations of the complex spherical harmonics $Y_l^m$ \cite{Helgaker_2000}. The rotation of $Y_l^m$ are characterized by the Wigner D-matrix \cite{Wigner_1931, Bradley_1972}
\begin{align}
    \alpha Y_l^{m} = Y_l^{m'} D_{m'm}^l(\alpha) \, ,
\end{align}
where $\alpha$ is a proper rotation, i.e. the rotation matrix has positive determinant. For an improper rotation $\alpha'$, $D_{mm'}^l(\alpha') = (-1)^{l} D_{mm'}^l(\alpha)$, with $\alpha$ the corresponding proper rotation. 
In general, these matrices determine how different magnetic quantum number $m$ mix for a given $l$.
Explicit expressions for the Wigner D-matrix are given in textbooks \cite{Bradley_1972}.
The matrix $\bO^{\bk}(\hat{\alpha})$ forms a representation of the symmetry group in orbital space.

\subsection{Factor groups and projective representations}

Symmetry operations can be used to find a symmetry-adapted wave-function basis that block-diagonalizes quantities, which each block belonging to a different irreducible representation of the symmetry group. The case of abelian groups is particularly simple, since all irreducible representations are one-dimensional. In the case of the translation group, the resulting blocks are typically labeled by their momentum index and lead to the usual momentum-space formalism.

Space-group operations beyond translations may lead to further block structure at high symmetry points, which can be exploited to efficiently perform calculations.
To obtain the unitary transformation matrix $\bU^{\bk}$ for  block diagonalization at each $k$ point, we introduce the concepts of the little co-group, the little group, and the projective representation.

For a momentum $\bk$, the little co-group $\overline{G}_{\bk}$ is defined as the point symmetry subgroup of the isogonal of the space group that leaves $\bk$ invariant in the reciprocal space, while the little group ${G}_{\bk}$ is the subgroup of elements whose rotation part $\alpha$ belongs to the little co-group $\overline{G}_{\bk}$ \cite{Bradley_1972}.
Finding $\bU^{\bk}$ corresponds to finding the projective irreducible representations of ${G}_{\bk}$ at each $k$ point \cite{Lax_1974, Dovesi_1998}. 

A projective representation of a finite group $G$ is a set of matrices that fulfills the condition \cite{Lax_1974}
\begin{align}
    \bD(\alpha) \cdot \bD(\beta) = \lambda(\alpha, \beta) \bD(\alpha \beta) \, , \, \, \, \, \alpha, \beta \in G \, ,
\end{align}
where $\lambda(\alpha, \beta)$ is a complex unitary factor that satisfies
\begin{align}
    \lambda(\alpha, \beta \gamma) \lambda(\beta, \gamma) = \lambda(\alpha\beta, \gamma) \lambda(\alpha, \beta) \, .
\end{align}
For each $k$ point, the projective representations and factors can be computed from the representation matrices \cite{Lax_1974, Dovesi_1998}
\begin{align}
    \bD^{\bk}(\alpha) = \text{exp}[i\bk \cdot v(\alpha)] \bO^{\bk} (\hat{\alpha}) \, , \label{eq:projrep}
\end{align}
\begin{align}
    \lambda^{\bk}(\alpha, \beta) = \text{exp} \{i\bk \cdot [v(\beta) - \alpha v(\beta)]\} \, . \label{eq:factor}
\end{align}
For a symmorphic group, the translation vectors $v(\alpha)$ for all symmetry operations  are 0, so that $D^{\bk}(\alpha) = O^{\bk}(\hat{\alpha})$, $\lambda^{\bk} = 1$.

\subsection{Symmetry adaptation and block diagonalization}

The standard formalism of group theory proceeds to decompose a representation into its irreducible components by applying successive projections onto basis functions belonging to a particular representation \cite{Dresselhaus_2008}. This formalism, while convenient for analytic manipulation, requires detailed knowledge of the representations beyond what is available in character tables, and is therefore somewhat unwieldy in its application to general problems with arbitrary symmetries.
A very elegant alternative to the projection was pioneered in this context by Dovesi and collaborators \cite{Dovesi_1986,Dovesi_1998} and relies on a simultaneous diagonalization of Dirac characters instead of a projection. This is the formalism we follow here.

 
We first introduce the concepts of conjugacy classes and Dirac characters.
The conjugacy classes $\mathit{C}$ of a group $G$ are defined as the subsets of elements that satisfy the relation
\begin{align}
    \alpha, \gamma \in \mathit{C}  \, \, \Longleftrightarrow \, \, \exists \beta \in \mathit{G}, ~ s.t. ~ \beta \gamma = \alpha \beta \, .
\end{align}
The Dirac character $\Omega_c$ of a conjugacy class $C$ is defined as 
\begin{align}
    \Omega_c 
    &= \frac{n_c}{h} \sum_{\beta \in G} \bD(\alpha) \cdot \bD(\gamma) \cdot \bD(\beta)^{-1} \nonumber \\
    &= \sum_{\alpha \in C} \lambda(\beta_{\alpha}, \gamma) \lambda(\alpha, \beta_{\alpha})^{*} \bD(\alpha) \, ,
    \label{eq:dirac}
\end{align}
with $\alpha, \gamma \in C$, $\beta_{\alpha}$ any group element such that $\beta_{\alpha} \gamma = \alpha \beta_{\alpha}$.
Importantly, all Dirac characters commute with the representation of
all group elements and with each other \cite{Lax_1974}. A simultaneous diagonalization of all Dirac characters results in a unitary transform that decomposes any basis set into subsets associated to each irreducible representation \cite{Dovesi_1998}. 

A numerical algorithm proceeds as follows: 
Given the projected orbital representation, all Dirac characters are formed using Eq.~\ref{eq:dirac}, and a character is considered as `relevant' if it is non-zero.
The set of eigenvectors that simultaneously diagonalize all  relevant Dirac characters can be obtained by successively diagonalizing the degenerated eigen subspace of all matrices following the methods presented in Ref.~\cite{Nobel_1977}.

As the eigenvalues of $\Omega_C$ are related to the characters of the irreducible representations, the block sizes corresponding to different irreducible representations can be obtained by finding the shortest common constant pieces in all diagonalized Dirac characters.
This simple simultaneous diagonalization approach may run into instability issues, and more precise algorithms such as the ones presented in Ref.~\cite{Angelika_1998} are necessary for large orbital representation matrices.
The full procedure of finding the common eigen space of the Dirac characters and identifying the corresponding irreducible representations in a computationally stable way is presented in Ref.~\cite{Dovesi_1998} which makes use of the Hermitian components of the Dirac characters.

\everymath{\displaystyle}
\begin{algorithm}
\setstretch{1.35}
    \caption{Transformation matrix}
    \label{alg:transform}
    \begin{algorithmic}[1] 
      \State Construct representation matrices $\bO^{\bk} (\hat{\alpha})$ for all $k$ points and symmetry operations.
      \algorithmiccomment{Eq.~\ref{eq:basis_rot}}
      \State Compute projective representation matrices $\bD^{\bk} ({\alpha})$ and factors $\lambda^{\bk}(\alpha, \beta)$.
      \algorithmiccomment{Eq.~\ref{eq:factor}}
      \State Compute Dirac characters $\Omega_c^{\bk}$ of the little group ${G}_{\bk}$ for all $k$ points.
      \algorithmiccomment{Eq.~\ref{eq:dirac}}
      \State Perform simultaneous diagonalization of the Dirac characters of all relevant classes. The common eigenvectors are the  transformation matrices $\mathbf{U}^{\bk}$ required to perform block diagonalization.
      \\
      \Return $\mathbf{O}^{\bk}(\hat{\alpha})$, $\mathbf{U}^{\bk}$.
    \end{algorithmic}
\end{algorithm}
We further discuss the consequence of including the generalized time-reversal operator $\bar{\kappa} = \phi \kappa$, in which $\kappa$ is the complex conjugation operator and $\phi$ is any point operator that gives the relation $\phi\bk = -\bk + \bK$ for a given $k$ point \cite{Lax_1974}.
The time-reversal symmetry indicates that the matrix representation of any time-reversal independent operator at $\phi\bk$ should be the complex conjugate of its matrix representation at $\bk$.
When $\bk = -\bk + \bK$, the matrix representation is real. In the symmetry adaptation procedure, if a given complex vector $\psi$ belongs to an irreducible set of the group, the pair of real vectors $\psi^{+} = \frac{1}{2}(\psi + \psi^{*})$ and $\psi^{-} = \frac{1}{2i}(\psi - \psi^{*})$ belongs to the same irreducible co-representation. Therefore, it is possible to construct a real symmetry adapted basis set from the complex eigen vectors by simultaneous diagonalization \cite{Dovesi_1998}. 

The procedure we use to obtain $\bO^{\bk}(\hat{\alpha})$, $\bU^{\bk}$ is summarized in Algorithm \ref{alg:transform}.

\section{Application to diagrammatic calculations}

\subsection{Finite-temperature GW for periodic systems} \label{sec:gw}

In finite-temperature many-body theory, the electronic properties of a system are characterized by the single-particle Green's function
\begin{align}
    G^{\bk}_{ij, \, \sigma}(\tau) = -\langle \T c_{i\sigma}^{\bk}(\tau) c_{j\sigma}^{\bk \dagger}(0) \rangle \, ,
\end{align}
where $\beta = \frac{1}{k_B T}$ is the inverse temperature, $\tau \in [0, \beta]$ is the imaginary time, and $\mathcal{T}$ the time ordering operator \cite{Abrikosov_1965,Negele_1988}.
The interacting Green's function is related to the non-interacting Green's function via the Dyson equation
\begin{align}
    [\bG^{\bk}(i\omega_n)]^{-1} 
    &= [\bG_0^{\bk}(i\omega_n)]^{-1}  - \bSigma^{\bk}(i\omega_n) \nonumber \\
    &= (i\omega_n + \mu) \bS^{\bk} - \bH^{\bk}_0 - \bSigma^{\bk}(i\omega_n) \, , \label{eq:dyson}
\end{align}
in which $\mu$ is the chemical potential, $\bS^{\bk}$ is the overlap matrix of the basis functions
\begin{align}
    S^{\bk}_{ij} = \int_{\Omega} d\mathbf{r} g^{\bk*}_{i}(\mathbf{r})g^{\bk}_{j}(\mathbf{r}) \, ,
\label{Eq:scalar_S}
\end{align}
and $\bG^{\bk}(i\omega_n)$ is the Matsubara frequency Green's function
\begin{align}
    \bG^{\bk}(i\omega_n) = \int_0^{\beta} d\tau \, \bG^{\bk}(\tau) e^{i\omega_n \tau} \, ,
\end{align}
with $\omega_n = (2n+1)\pi / \beta$ fermionic Matsubara frequencies. The self-energy $\bSigma^{\bk}(i\omega_n)$ is a function of the full Green's function $\bSigma \equiv \bSigma[\bG]$,
which can be split into static and dynamic parts
\begin{align}
    \bSigma[\bG](i\omega_n) = \bSigma^\text{(HF)}[\bG] + \tilde{\bSigma}[\bG] (i\omega_n) \, . \label{eq:sigma_split}
\end{align}
In many-body simulations of realistic systems, the self-energy is usually approximated by low-order diagrams. The self-consistent GW method is a common choice; it approximates the dynamical part of the self-energy as the sum of an infinite series of RPA-like “bubble” diagrams \cite{Hedin_1965}. The explicit equation for the GW self-energy reads
\begin{align}
    (\tilde{\Sigma}^{\text{GW}})^{\bk}_{i\sigma,j\sigma}&(\tau) = \frac{-1}{N_{k}}\sum_{\bq}\sum_{ab}\sum_{QQ'} \label{eq:sigma} \\
    &G^{\bk-\bq}_{a\sigma,b\sigma}(\tau)V^{\bk,\bk-\bq}_{\ i \ \ a}(Q) {\bP}^{\bq}_{QQ'}(\tau)V^{\bk-\bq,\bk}_{\ \ b \ \ \ j}(Q') \, . \nonumber 
\end{align}
In Eq.~\ref{eq:sigma}, $\bV^{\bk_i \bk_j}$ decomposes the Coulomb interaction as
\begin{align}
    U^{\bk_{i}\bk_{j}\bk_{k}\bk_{l}}_{\ i\ j\ \ k\ l} = \sum_{Q}V^{\bk_{i}\bk_{j}}_{\ i\ j}(Q)V^{\bk_{k}\bk_{l}}_{\ k\ l}(Q) \, ,
    \label{Eq:U_decompose}
\end{align}
with $Q$ an auxiliary basis index.
We use density fitted interactions ~\cite{DF_Werner2003,DF_Ren2012,Sun_2017} in our calculations. These tensors can be expressed as
\begin{align}
    V^{\bk_{i}\bk_{j}}_{\ i \ j}(Q) = \sum_{P} (\mathbf{J}^{\bq})^{-1/2}_{QP} C^{\bk_{i}\bk_{j}\bq}_{\ i\ j}(P) \, ,
\label{Eq:VijQ}
\end{align}
with the relations $\bq = \bk_j - \bk_i$ and $\bJ^{-\bq} = \bJ^{\bq *}$, where $J^{\bq}$ is the two-center integral related to the auxiliary basis. The explicit expressions of the density fitting integrals can be found in Ref.~\cite{Sun_2017}.
The auxiliary function $\bP^{\bq}(\tau)$ is defined through the geometric series
\begin{subequations}
\begin{align}
    {\mathbf{P}}^{\mathbf{q}}(i\Omega_{n}) = [\mathbf{I} - {\mathbf{P}}^{\mathbf{q}}_{0}(i\Omega_{n})]^{-1}{\mathbf{P}}^{\mathbf{q}}_{0}(i\Omega_{n}) \, ,
\end{align}
\begin{align}
    {P}^{\bq}_{QQ}(\tau) = \frac{1}{\beta}\sum_{n}{P}^{\mathbf{q}}_{QQ'}(i\Omega_{n})e^{-i\Omega_{n}\tau} \, ,
\end{align}
\label{eq:P02P}
\end{subequations}
where $\Omega_{n} = 2n\pi / \beta$ are bosonic Matsubara frequencies. ${\bP}^{\bq}_{0}(\tau)$ is an auxiliary function related to the `bare bubble', which is defined as
\begin{align}
    {P}^{\bq}_{0,QQ'}(\tau)& = \frac{-1}{N_{k}}\sum_{\bk}\sum_{\sigma\sigma'}\sum_{abcd}V^{\bk,\bk+\bq}_{\ d \ a}(Q) \label{eq:P0} \\
    &\times G^{\bk}_{c\sigma',d\sigma}(-\tau)G^{\bk+\bq}_{a\sigma \ b\sigma'}(\tau)V^{\bk+\bq,\bk}_{\ \ b\  \ \ c}(Q') \, . \nonumber
\end{align}
The static self-energy is the Hartree-Fock self-energy obtained with the interacting density matrix,
\begin{subequations}
\begin{align}
    &\bSigma^\text{(HF)} = \bJ^{\bk} + \bK^{\mathbf{k}} \, , 
    \\
    &J^{\mathbf{k}}_{i\sigma,j\sigma} = \frac{1}{N_{k}}\sum_{\mathbf{k}'}\sum_{\sigma_{1}}\sum_{ab}\sum_{Q}V^{\mathbf{k}\mathbf{k}}_{\ ij}(Q)\gamma^{\mathbf{k}'}_{a\sigma_{1},b\sigma_{1}}V^{\mathbf{k}'\mathbf{k}'}_{\ b\ a}(Q) \, ,
    \\
    &K^{\mathbf{k}}_{i\sigma,j\sigma'} = -\frac{1}{N_{k}}\sum_{\mathbf{k}'}\sum_{ab}\sum_{Q}V^{\mathbf{k}\mathbf{k}'}_{\ i\ a}(Q)\gamma^{\mathbf{k}'}_{a\sigma,b\sigma'}V^{\mathbf{k}'\mathbf{k}}_{\ b\ j}(Q) \, ,
\end{align}
\label{eq:sigma_hf}
\end{subequations}
where $\mathbf{\gamma}^{\bk} = \bG^{\bk}(\tau=0^{-})$ is the density matrix.
The self-consistent GW method solves Eqs.~\ref{eq:dyson},\ref{eq:sigma_split},\ref{eq:sigma},\ref{eq:sigma_hf} in a self-consistent manner for a system with a fixed number of electrons $N_e = \frac{1}{N_k}\sum_{\bk} \text{Tr}[\mathbf{\gamma}^{\bk} \bS^{\bk}]$.
For detailed derivations and a motivation of the formalism used here see Refs.~\cite{Yeh_22} and \cite{GREEN}.

\subsection{Rotation to the irreducible Brillouin zone}

To obtain the  solution of the self-consistent many-body equations, it is necessary to compute the Green's function and self-energies at  all momentum sampling points within the first Brillouin zone. These evaluations can be simplified with symmetry considerations.

Using rotation matrices that rotate orbitals between different $k$ points is a common way of reducing computational complexity.  This formalism is implemented in essentially all Hartree Fock and DFT codes, including in Refs.~\cite{Dovesi_1998,aims_2022}.

The transformation of the matrix representation of an operator $\hat{X}$ that is invariant under a given symmetry operation $\hat{\alpha}$ reads
\begin{align}
    &\mathbf{X}^{\tilde{\bk}} = \mathbf{O}^{\bk}(\hat{\alpha}) \, \mathbf{X}^{\bk} \, \mathbf{O}^{\bk \dagger}(\hat{\alpha}) \, , \label{eq:mat_rot}
\end{align}
where $X_{ij}^{{\bk}} = \langle g_i^{{\bk}} | \hat{X} | g_j^{{\bk}} \rangle$ and $\alpha\bk = \tilde{\bk} + \bK$.
In the self-consistent GW method outlined in Section~\ref{sec:gw}, the transformations of the overlap matrix $\bS$, bare Hamiltonian $\bH_0$, Green's function $\bG$ and self-energy $\bSigma$ follow Eq.~\ref{eq:mat_rot} with $\mathbf{O}^{\bk}(\hat{\alpha}) \equiv \bO^{\bk}_\text{ao}(\hat{\alpha})$ the representation matrix in atomic orbital space.

Besides the calculation of matrix-valued functions, the calculation and storage of the tensor-valued decomposed interaction tensor $\bV^{\bk_i \bk_j}$ also serves as a potential bottleneck in many-body calculations.
The transformation of $\bV^{\bk_i \bk_j}$ can be derived from its definition Eq.~\ref{Eq:VijQ}.
The two-index matrix $\bJ^{\bq}$ transforms as Eq.~\ref{eq:mat_rot} with $\mathbf{O}^{\bk}(\hat{\alpha}) \equiv \bO^{\bq}_\text{aux}(\hat{\alpha})$ the representation matrix in auxiliary basis space, and the three-index tensor transforms as
\begin{align}
    \bC^{\tilde{\bk}_{i} \tilde{\bk}_{j} \tilde{\bq}}
    =
    \bO^{\bq}(\hat{\alpha}) \bO^{\bk_i}(\hat{\alpha}) \bC^{{\bk}_{i} {\bk}_{j} {\bq}} O^{{\bk}_j \dagger} (\hat{\alpha}) \, .
\end{align}
Threrfore, the transformation of the decomposed interaction can be written as
\begin{subequations}
\begin{align}
    &\bV^{\tilde{\bk}_{i} \tilde{\bk}_{j}}
    =
    (\bJ^{\tilde{\bq}})^{-1/2} {\bO}^{{\bq}}(\hat{\alpha}) (\bJ^{\bq})^{1/2}
    \bO^{\bk_i}(\hat{\alpha}) \bV^{{\bk}_{i} {\bk}_{j}} O^{{\bk}_j \dagger} (\hat{\alpha}) \nonumber \\
    &\phantom{\bV^{\tilde{\bk}_{i} \tilde{\bk}_{j}}}= 
    \overline{\bO}^{\bq}(\hat{\alpha}) \bO^{\bk_i}(\hat{\alpha}) \bV^{{\bk}_{i} {\bk}_{j}} O^{{\bk}_j \dagger} (\hat{\alpha})
    \, , 
\\
    &\overline{\bO}^{{\bq}}(\hat{\alpha})
    = (\bJ^{\tilde{\bq}})^{-1/2} {\bO}^{{\bq}}(\hat{\alpha}) (\bJ^{\bq})^{1/2} \, .
\end{align}
\end{subequations}
Given the transformation of $\bV^{\bk_i \bk_j}$, the transformations of $\bP_0^{\bq}$ and $\bP^{\bq}$ can be derived from their definitions in Eqs.~\ref{eq:P02P}, \ref{eq:P0} since the auxiliary indices contained in these two objects originated from those of $\bV^{\bk_i \bk_j}$.
The rotation of $\bP_0^{\bq}$ and $\bP^{\bq}$ reads
\begin{align}
    &\bP^{\tilde{\bq}} = \overline{\bO}^{\bq}(\hat{\alpha}) \, \bP^{\bq} \, \overline{\bO}^{\bq \dagger}(\hat{\alpha}) \, .
\label{eq:prot}
\end{align}

In momentum space, sets of $k$ points connected by symmetry operations are called stars \cite{Bradley_1972}. 
From each star, only one representative is chosen; properties at other elements of the star are regenerated by transformation. The set of representatives of the star delineates the irreducible wedge or irreducible Brillouin zone (IBZ). 
This formulation reduces the calculations from the full Brillouin zone into the irreducible wedge. 
All two-index matrices (which are dependent on only a single momentum index) can be computed within the IBZ and subsequently transformed to other $k$ points using the equations outlined in Eqs.~\ref{eq:mat_rot}, \ref{eq:prot}.

In tensors with more than two indices, such as three-index tensors which depend on two momenta, it is feasible to choose one momentum index confined within the IBZ while the remaining indices span the full Brillouin zone; the choice of the index within the IBZ is arbitrary.

\subsection{Block diagonalization} \label{sec:block}
In systems where a large number of momentum points lie on the surface of the IBZ, it is advantageous to further reduce the computational costs of all the tensor contractions and linear solvers by block diagonalization.
The matrix at each $k$ point can be transformed into a block diagonal form with the unitary transformation
\begin{align}
    {\bX}^{{\bk}}_{\text{block}} = \bU^{{\bk} \dagger} \bX^{{\bk}} \mathbf{U}^{{\bk}} \, .
    \label{eq:mat_blk}
\end{align}
To obtain the corresponding block form of the decomposed interaction, we consider the three-index tensor and two-index matrix in Eq.~\ref{Eq:VijQ} separately.
The three-index tensor transforms as
\begin{align}
    {\bC}^{\bk_i \bk_j \bq}_{\text{block}}
    = \bU^{\bk_i \dagger} \bU^{{\bq} \dagger} C^{\bk_i \bk_j \bq} \bU^{\bk_j} \, , \label{eq:C_blk}
\end{align}
and the two-index tensor $\bJ^{\bq}$ can be block diagonalized using the transformation in Eq.~\ref{eq:mat_blk}.
To ensure the block diagonalized structure of $\bP_0^{\bq}$ and $\bP^{\bq}$, we perform block diagonalization of $\bJ^{\bq}$ first and then decompose each block separately to obtain $(\bJ^{\bq})^{-1/2}$ when constructing the decomposed interaction in Eq.~\ref{Eq:VijQ}.
The rotation between block diagonalized matrices at different $k$ points can be achieved using the block transformed rotation matrix
\begin{align}
    {\bO}^{{\bk}}_{\text{block}}(\hat{\alpha}) = \bU^{\tilde{\bk} \dagger} \bO^{{\bk}}(\hat{\alpha}) \bU^{{\bk}} \, .
\end{align}
Only non-zero blocks are considered in the evaluation of diagrammatic equations.

\section{Results for many-body calculations}

\subsection{Floating point operations}

\begin{center}
\begin{table}[tbh]
\begin{tabular}{ccccc}\\
 \toprule
 Compound & Group & Symmorphic & $n_{ao}$ & $n_{aux}$ \\ [0.5ex] 
 \midrule
 Si & 227 & False & 26 & 124 \\
 \midrule
 BN & 194 & False & 52 & 192 \\
 \midrule
 AlP & 216 & True & 26 & 124 \\
 \midrule
 GaAs & 216 & True & 38 & 254 \\
 \midrule
 \bottomrule
\end{tabular} \\[0.2cm]
\caption{Symmetry groups and basis set sizes of four example systems.}
\label{tab:pbc}
\end{table}
\end{center}
%
\begin{center}
\begin{table}[tbh]
\begin{tabularx}{0.48\textwidth}{cccccc}\\
 \toprule
 & & & \multicolumn{3}{@{}c@{}}{Floating-point operations}  \\ [0.5ex] 
 \midrule
 & $n_k$ & $n_{ik}$ & Full & Rotation & Block Diag\\
 \midrule
 \multicolumn{6}{@{}l@{}}{Si} \\
 \midrule
 & 1 & 1 & $1.31\times 10^{10}$ & $1.31\times 10^{10}$ & $1.50\times 10^{9}$ \\
 \midrule
 & 2 & 3 & $1.73\times 10^{12}$ & $1.01\times 10^{12}$ & $2.24\times 10^{11}$ \\
 \midrule
 & 4 & 8 & $1.10\times 10^{14}$ & $2.13\times 10^{13}$ & $8.55\times 10^{12}$ \\
 \midrule
 & 6 & 16 & $1.25\times 10^{15}$ & $1.43\times 10^{14}$ & $6.87\times 10^{13}$ \\
 \midrule
 \multicolumn{6}{@{}l@{}}{BN} \\
 \midrule
 & 1 & 1 & $1.12\times 10^{11}$ & $1.12\times 10^{11}$ & $4.24\times 10^{9}$ \\
 \midrule
 & 2 & 4 & $2.72\times 10^{13}$ & $1.37\times 10^{13}$ & $3.99\times 10^{12}$ \\
 \midrule
 & 4 & 12 & $1.74\times 10^{15}$ & $3.27\times 10^{14}$ & $1.30\times 10^{14}$ \\
 \midrule
 & 6 & 28 & $1.98\times 10^{16}$ & $2.57\times 10^{15}$ & $1.33\times 10^{15}$ \\
 \midrule
 \multicolumn{6}{@{}l@{}}{AlP} \\
 \midrule
 & 1 & 1 & $1.31\times 10^{10}$ & $1.31\times 10^{10}$ & $5.31\times 10^{9}$ \\
 \midrule
 & 2 & 3 & $2.65\times 10^{12}$ & $1.01\times 10^{12}$ & $4.00\times 10^{11}$ \\
 \midrule
 & 4 & 10 & $1.68\times 10^{14}$ & $2.65\times 10^{13}$ & $1.32\times 10^{13}$ \\
 \midrule
 & 6 & 22 & $1.92\times 10^{15}$ & $1.96\times 10^{14}$ & $1.15\times 10^{14}$ \\
 \midrule
 \multicolumn{6}{@{}l@{}}{GaAs} \\
 \midrule
 & 1 & 1 & $9.59\times 10^{10}$ & $9.59\times 10^{10}$ & $3.32\times 10^{10}$ \\
 \midrule
 & 2 & 3 & $2.15\times 10^{13}$ & $8.21\times 10^{12}$ & $3.14\times 10^{12}$ \\
 \midrule
 & 4 & 10 & $1.37\times 10^{15}$ & $2.16\times 10^{14}$ & $9.40\times 10^{13}$ \\
 \midrule
 & 6 & 22 & $1.56\times 10^{16}$ & $1.59\times 10^{15}$ & $9.11\times 10^{14}$ \\
 \bottomrule
\end{tabularx} \\[0.2cm]
\caption{Flop counts of GW calculations. $n_k$ is the number of $k$ points in each direction, and $n_{ik}$ is the total number of $k$ points in irreducible wedge after symmetry considerations.}
\label{tab:flops}
\end{table}
\end{center}

The computationally expensive part in the self-consistent GW calculation is the evaluation of the self-energy (Eqs.~\ref{eq:P0}, \ref{eq:P02P}, \ref{eq:sigma}). 
As an illustrative example, we choose 4 compounds consisting of elements in group III, IV, V to show the reduction of the total number of floating-point operations with symmetry adaptation. 
See Table~\ref{tab:pbc} for the detail structure information of all considered compounds. All the space group analysis are performed using the \texttt{spglib} software \cite{spglib_v1}.
Table~\ref{tab:flops} summarizes the results on a momentum mesh of size  $n_k \times n_k \times n_k$ centered at the $\Gamma$ point with $n_k = 1, 2, 4, 6$. 
All calculations are performed using the \texttt{gthdzvp} basis and the \texttt{def2-svp-ri} auxiliary basis
with 114 imaginary time points and 103 bosonic frequency points on an intermediate representation (IR) grid \cite{Shinaoka_2017,Li_2020}.
We use the \pySCF package \cite{Sun_2018} to generate the integrals in Bloch wave bases and block-diagonalize them as described in section ~\ref{sec:block} before starting the self-consistent GW calculation.
When $n_k=1,$ the number of floating-point operations is computed using the expression for type \texttt{double} instead of \texttt{complex} \texttt{double} as all matrices and tensors are real in this case.

As shown in the table, as the total number of $k$ points increases, the inter-$k$ point rotations significantly reduce the overall number of floating-point operations, while block diagonalization is particularly advantageous in systems with few momentum points.
This is because with an increasing number of $k$ points, fewer points lie on the surface of the IBZ and
the number of operations in the little group $G_\bk$ of most of the considered $k$ points is low. Therefore, there will not be many zero blocks in most of the matrices, and the computational advantage of doing block diagonalization decreases~\cite{Dovesi_1998_1}.
In general, employing both rotation and block diagonalization results in roughly one order of magnitude fewer floating-point operations across all the examples considered.

\subsection{GPU acceleration}
Symmetry adaptation offers the possibility to expose otherwise inaccessible opportunities for parallelism. To facilitate the expensive self-energy evaluations, we design an acceleration scheme leveraging both MPI and CUDA parallelization.
Algorithm \ref{alg:solver} shows the pseudo code of a GPU accelerated GW self-energy solver. As shown, the computationally expensive evaluations in Eqs.~\ref{eq:sigma}, \ref{eq:P0} are performed on GPU, and Eq.~\ref{eq:P02P} is evaluated on CPU using MPI parallelization.

The parallelization design in Ref.~\cite{Yeh_22} assigns the calculation of each $\bq$ point to one GPU card. For calculations without symmetry transformations, this design avoid storing the full auxiliary tensor $\bP^{\bq}(\tau)$ of all $\bq$ points, but also limits the maximum number of GPU card used to the number of $\bq$ points.

In contrast, in the scheme introduced here for symmetry-adapted calculations, Eq.~\ref{eq:P0} is evaluated for $\bq$ in IBZ and Eq.~\ref{eq:sigma} is evaluated for $\bk$ in IBZ, with both $\bP^{\bq}(\tau)$ and $\tilde{\bSigma}^{\bk}(\tau)$ in block diagonal form.
Besides reducing the flop count, this transformation also reduces the memory required for storing these quantities.
We treat $\bq$ and $\bk$ in Eqs.~\ref{eq:P0} and \ref{eq:sigma} on equal footing by generating $(\bk_1, \bk_2)$ pairs with $\bk_1$ in IBZ and $\bk_2$ in full BZ.
Multiple asynchronous streams (parallel threads) are created on each GPU card and 
the calculation associated with each $(\bk_1, \bk_2)$ pair is assigned to these streams. Asynchronous stream handling allows for overlap between dense tensor contractions of all $(\bq, \bk)$ pairs.
For each $(\bq, \bk)$ pair, the loop over $n_\tau$ are performed in a batched way using strided batched $\zgemm$ cuBlas calls to reach peak performance.
See Algorithm \ref{alg:P0} and \ref{alg:sigma} for the pseudo code of GPU calculations.

As each GPU card usually has less memory than a CPU node, one critical aspect of GPU acceleration is the balance between memory allocation on GPU and memory transfer between GPU and CPU. 
In our design, 
the Green's function $\mathbf{G}^{\bk}(\tau)$, self-energy $\tilde{\bSigma}^{\bk}(\tau)$, auxiliary functions $\bP^{\bq}_{0}(\tau)$, $\bP^{\bq}(\tau)$, and rotation matrices $\bO^{\bk}_\text{ao}$, $\bO^{\bk}_\text{aux}$ for all required momenta are stored in the shared memory of each GPU card to avoid frequent copy of memory between GPU and CPU.
The decomposed interaction $\bV^{\bk_1 \bk_2}$ for all $(\bk_1, \bk_2)$ pairs are usually too large to store on the GPU card, even with symmetry adaptation. Therefore, we store it either on disk or in the shared memory of each CPU node and copy data to the GPU on demand.
As all streams created on the GPU cards may run in parallel, each stream stores a local copy of $\tilde{\bSigma}^{\bk,\bq}(\tau)$, $\bP_0^{\bq,\bk}(\tau)$ and $\bV^{\bk_1 \bk_2}$ for a single $(\bq, \bk)$ pair, such that $\tilde{\bSigma}^{\bk}(\tau) = \sum_{\bq} \tilde{\bSigma}^{\bk,\bq}(\tau)$, $\bP_0^{\bq}(\tau) = \sum_{\bk} \bP_0^{\bq,\bk}(\tau)$.

\begin{figure}[tbh]
    \centering
    \includegraphics[scale=1]{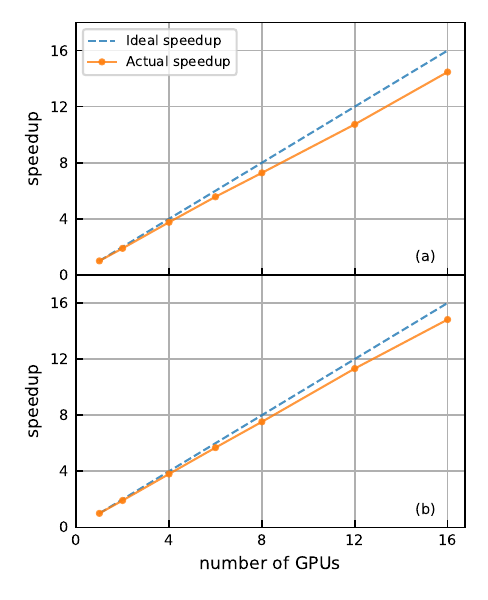}
    \caption{Profile of the GPU kernels with up to 16 GPU cards. (a): Algorithm \ref{alg:P0}. (b): Algorithm \ref{alg:sigma}.
    }
    \label{fig:speedup}
\end{figure}

Fig.~\ref{fig:speedup} shows a profile of Algorithms \ref{alg:P0}, \ref{alg:sigma} in our implementation. The profile is obtained on a cluster with each node containing eight V100 NVIDIA GPU cards. We use Si on a momentum mesh of size $4 \times 4 \times 4$  centered at $\Gamma$ point as the test system. The number of orbitals, auxiliary basis, and irreducible $k$ points can be found in Tables \ref{tab:pbc}, \ref{tab:flops}. Sixteen asynchronous streams are created on each GPU card.
As shown in the figure, the GPU kernels of $\bP_0$ and $\tilde{\bSigma}$ contractions shows almost ideal speedup.

\everymath{\displaystyle}
\begin{algorithm}
\setstretch{1.35}
    \caption{GPU accelerated GW solver}
    \label{alg:solver}
    \begin{algorithmic}[1] 
      \State
      \textrm{Copy} $\mathbf{G}^{\bk}(\tau)$, $\bO^{\bk}_\text{ao}$ \textrm{to GPU.}
      \State
      \textrm{Compute} $\mathbf{G}^{\bk}(\tau)$ \textrm{in full BZ.} \algorithmiccomment{Eq.~\ref{eq:mat_rot}}
      \State
      \textrm{Compute} $\mathbf{P}_0^{\bq}(\tau)$ \textrm{on GPU.} \algorithmiccomment{Algorithm \ref{alg:P0}}
      \State
      \textrm{Collect} $\mathbf{P}_0^{\bq}(\tau)$ \textrm{to CPU.} 
      \State
      \textrm{Solve} $\mathbf{P}^{\bq}(\tau)$ \textrm{with MPI parallelization.}
      \algorithmiccomment{Eq.~\ref{eq:P02P}}
      \State
      \textrm{Copy} $\mathbf{P}^{\bq}(\tau)$, $\overline{\bO}^{\bq}_\text{aux}$ \textrm{to GPU.}
      \State
      \textrm{Compute} $\tilde{\bSigma}^{\bk}(\tau)$ \textrm{on GPU.}
      \algorithmiccomment{Algorithm \ref{alg:sigma}} 
      \State
      \textrm{Collect} $\tilde{\bSigma}^{\bk}(\tau)$ \textrm{to CPU.} 
      \\
      \Return $\tilde{\bSigma}^{\bk}(\tau)$.
    \end{algorithmic}
\end{algorithm}
\everymath{\displaystyle}
\begin{algorithm}
\setstretch{1.35}
    \caption{Auxiliary function calculation (Eq.~\ref{eq:P0})}
    \label{alg:P0}
    \begin{algorithmic}[1] 
      \For{each $(\bq, \bk)$ pair}
      \State Assign calculation to an idle stream.
      \State Copy $\bV^{\bk,\bk+\bq}$ to GPU.
      \For{$\tau = 0$ to $n_{\tau}/2$ with step $n_\text{batch}$}
      \For{each non-zero block in $\bV^{\bk,\bk+\bq}$}
      \State $\bX_1(\tau) \leftarrow \bV^{\bk,\bk+\bq} \bG^{\bk}(-\tau)$
      \State $\bX_2(\tau) \leftarrow \bV^{\bk+\bq,\bk} \bG^{\bk+\bq}(\tau)$
      \State $\bP_0^{\bq,\bk}(\tau) \leftarrow \bX_1(\tau) \bX_2(\tau)$
      \EndFor
      \State Copy $\bP_0^{\bq,\bk}(\tau)$ to ${\bP}_0^{\bq}(\tau)$.
      \EndFor
      \EndFor \\
      \Return  ${\bP}_0^{\bq}(\tau)$.
    \end{algorithmic}
\end{algorithm}
\everymath{\displaystyle}
\begin{algorithm}
\setstretch{1.35}
    \caption{GW self-energy calculation (Eq.~\ref{eq:sigma})}
    \label{alg:sigma}
    \begin{algorithmic}[1] 
      \For{each $(\bq, \bk)$ pair}
      \State Assign calculation to an idle stream.
      \State Copy $\bV^{\bk,\bk-\bq}$ to GPU.
      \For{$\tau = 0$ to $n_{\tau}$ with step $n_\text{batch}$}
      \For{each non-zero block in $\bV^{\bk,\bk-\bq}$}
      \State $\bX_1(\tau) \leftarrow \bV^{\bk,\bk-\bq} \bG^{\bk-\bq}(\tau)$
      \State Compute $\bP^{\bq}(\tau)$ from $\bP^{\tilde{\bq}}(\tau)$.
      \algorithmiccomment{Eq.~\ref{eq:prot}}
      \State $\bX_2(\tau) \leftarrow \bX_1(\tau) \bP^{\bq}(\tau)$
      \State $\tilde{\bSigma}^{\bk,\bq}(\tau) \leftarrow \bX_2(\tau) \bV^{\bk-\bq, \bk}$
      \EndFor
      \State Copy $\tilde{\bSigma}^{\bk,\bq}(\tau)$ to $\tilde{\bSigma}^{\bk}(\tau)$.
      \EndFor
      \EndFor \\
      \Return $\tilde{\bSigma}^{\bk}(\tau)$.
    \end{algorithmic}
\end{algorithm}

\section{Conclusions}
In conclusion, we have presented results for the implementation, numerical cost, and scaling of the symmetry adaptation of solids within the a self-consistent many-body formalism. Our method relied on the simultaneous diagonalization of Dirac characters \cite{Lax_1974} as pioneered in the electronic structure context by \cite{Dovesi_1986,Dovesi_1998} and employed an efficient parallelization scheme on graphics accelerators.

Multiple future directions for exploration are evident. First, while this work focused on standard space group symmetries, generalizations of the framework to magnetic space groups or `Shubnikov' groups \cite{Tavger_1956} are straightforward and may be required for the efficient simulation of magnetic or relativistic systems. Second, the knowledge of symmetries and representations offers a straightforward connection to the interpretation of optical response functions, facilitating the interpretation and simulation of such experiments. Finally, more precise simulation methods may employ higher order terms in diagrammatic perturbation theory, such as `vertices', whose calculation will benefit both from the symmetry analysis and from the reduction in computational effort presented here.

\section{Acknowledgments}
EG was funded by the US National Science Foundation via grant NSF OAC 2310582. We thank Sergei Iskakov for useful comments.

\bibliographystyle{elsarticle-num} 
\bibliography{refs}

\end{document}